\documentclass[reprint,twocolumn,aps,prb,superscriptaddress]{revtex4-1}

\usepackage[T1]{fontenc}
\usepackage[utf8]{inputenc}
\usepackage{lmodern}
\usepackage{graphicx}
\usepackage{amsmath}
\usepackage{xcolor}
\usepackage{float}
\renewcommand{\vec}[1]{\mathbf{#1}}
\usepackage{hyperref}
\usepackage{amssymb}
\usepackage{amsfonts}
\usepackage{amsmath}
\usepackage{amsbsy}
\usepackage{natbib}
\usepackage{comment}

\usepackage{color}
\usepackage{float}

\usepackage{mathtools} 
\usepackage{graphicx}
\usepackage{xcolor}
\usepackage[T1]{fontenc}
\usepackage[utf8]{inputenc}
\usepackage{lmodern}
\usepackage{amsmath}
\usepackage{float}
\renewcommand{\vec}[1]{\mathbf{#1}}
\usepackage{hyperref}
\usepackage{amssymb}
\usepackage{amsfonts}
\usepackage{amsmath}
\usepackage{amsbsy}
\usepackage{natbib}
\usepackage{comment}

\usepackage{color}
\usepackage{float}
\usepackage{xr}

\begin{document}
\renewcommand{\vec}[1]{\mathbf{#1}}
\newcommand{\ii}{\mathrm{i}}

\title{Topological dynamical quantum phase transition in a quantum skyrmion phase}

\author{Vipin Vijayan}
\affiliation{Department of Physics, Indian Institute of Technology (Banaras Hindu University)}

\author{L. Chotorlishvili}
\affiliation{Department of Physics and Medical Engineering, Rzesz\'ow University of Technology, 35-959 Rzesz\'ow, Poland}

\author{A. Ernst\mbox{*}}
\affiliation{Max Planck Institute of Microstructure Physics, Weinberg 2, D-06120 Halle, Germany}
\affiliation{Institute for Theoretical Physics, Johannes Kepler University, Altenberger Stra\ss e 69, 4040 Linz, Austria}

\author{S.\,S.\,P.  Parkin}
\affiliation{Max Planck Institute of Microstructure Physics, Weinberg 2, D-06120 Halle, Germany}

\author{M.\,I. Katsnelson}
\affiliation{Radboud University, Institute for Molecules and Materials, Heyendaalseweg 135, 6525AJ Nijmegen, Netherlands}

\author{S.\,K. Mishra}
\affiliation{Department of Physics, Indian Institute of Technology (Banaras Hindu University)}

\date{\today}

\begin{abstract}
  Quantum skyrmionic phase is modelled in a 2D helical spin lattice.
  This topological skyrmionic phase retains its nature in a large
  parameter space before moving to a ferromagnetic phase. Next
  nearest-neighbour interaction improves the stability and it also causes a shift of the
  topological phase in the parameter space.  Nonanalytic behaviour of
  the rate function observed, when the system which is initially in a
  quantum skyrmion phase is quenched to a trivial quantum
  ferromagnetic phase, indicates a dynamical quantum phase
  transition. Dynamical quantum phase transition is absent when the
  system initially in a skyrmion phase is quenched to a helical phase.
\end{abstract}

\maketitle

Progresses made in the last decade in controlling matter at quantum
levels allow access to real-time dynamics of closed quantum
many-body systems realisable
\cite{2008RvMP...80..885B,2012NatPh...8..267B}. This progress lifted the border
between experimentally feasible physical reality and model
systems. Ultra-cold atoms in
optical lattices and trapped ions are examples in which such dynamical phenomena 
 were observed in real-time. Nowadays, we have full access to the real-time dynamics of
quantum many-body and finite systems, either isolated or coupled to
the Markovian or non-Markovian environment. The experiments with THz
pulses in solids
\cite{teraherts_1,teraherts_2,teraherts_3,teraherts_4}, high magnetic
field pulse experiments \cite{huge_mag}{\it etc.} are also
developments in recent past which can be aided by theoretical
understanding of dynamical properties of the corresponding quantum
systems especially the study of the evolution of the system after a
sudden change in its parameter or a quantum quench. When we talk about
changing the parameter of a system, the first thing that pops up in
our mind is the term phase transition. Phase transitions are the
points in the parameter space of a system around which a small change
in the control parameter manifests a drastic change to its
characteristics. In a classical/thermal phase transition, the thermal
fluctuations cause the destruction of long-range ordering and
facilitate the phase transition. But when we study the changes of the
parameters at zero temperature or ground states the characteristics of
the phase transition become purely quantum, because here the phase
transition is facilitated by quantum fluctuations instead of thermal
fluctuations. Such phase transitions are known as quantum phase
transitions. Equilibrium quantum phase transitions(EQPT) are studied
extensively but we have a lesser understanding of quantum systems out
of equilibrium. In order to theoretically aid the experimental
developments mentioned earlier we need to understand the dynamical
quantum phase transitions(DQPT). Recent experimental developments
identified a signature of dynamical behaviour after a quench in a
Haldane-like system \cite{2018NatPh..14..265F}. Experiments with
trapped ions were able to directly observe DQPTs
\cite{PhysRevLett.119.080501}. The theoretical inspiration for the
DQPT can be extracted from Lee-Yang theorem, Fisher zero's and
accompanying analysis. (For more details refer to \cite{Zvyagin_DQPT})

From Lee-Yang analysis  \cite{PhysRev.87.404, PhysRev.87.410} one can
arrive at the conclusion that for a partition function of external
fields (like magnetic field), which also depended on the temperature
$T$ as $Z(T)$, when zeros exist and have a positive real value then
each of those roots corresponds to nonanalyticity in the free energy,
{\it i.e.,} phase transition. Fisher extended this analysis
considering partition functions with complex temperature $z$ instead
of $T$, $Z(z)$. When Fisher's zeros of $Z(z)$ overlap with the real
axis it produces nonanalyticities or phase transitions, however, no
such overlaps is observed in the course of an EQPT. The real values of
Fisher's zeros correspond to a different kind of phase transition,
namely DQPT \cite{Zvyagin_DQPT, heyl2018dynamical}. Using these
analyses essence of DQPTs can be explained briefly as follows
\cite{PhysRevLett.110.135704,PhysRevB.102.174418,PhysRevLett.112.217204,2020PhRvB.101t5135G}:  

Short-lived non-equilibrium phase transitions accompanied with
a nonanalytic behaviour of physical quantities as a function of time
is a characteristic feature of DQPTs  \cite{PhysRevLett.110.135704}. To study this we utilise a quantity
$\mathcal{G}$ the Loschmidt amplitude as a function of time $t$, given as
\begin{eqnarray}\label{Loschmidt amplitude}
&& \mathcal{G}(t)=\langle\psi_0\vert e^{-i\hat H t}\vert\psi_0\rangle.
\label{eqn:Loschmidt amplitude}
\end{eqnarray}
In a sense, it plays the same role in the study of states out-of-equilibrium as 
the partition function $Z=Tr(e^{-\beta \hat H})$ in thermodynamic
equilibrium case. Here $\hat H$ is the Hamiltonian of the system and
$\beta$ is the inverse temperature and $\psi_0$ is the initial state
of the system. Here $Z$ can be seen as the kind of partition function
considered by Lee and Yang(with inverse temperature in place of
temperature, still the conclusions of the analysis holds). And in the
exponent term of $\mathcal{G}(t)$ if we take $(it)$ as the complex
temperature, $\mathcal{G}$ behaves like partition function of complex
temperature considered by Fisher. 

Another quantity of interest is the rate function of the return
probability, Eq.~\ref{eqn:prob. of return amplitude} (hereinafter
referred to as rate function) which is analogous to the thermodynamic
free energy. As discussed earlier during a phase transition the
thermodynamic free energy, $F=-\frac{ln(Z)}{\beta}$, turns out to be a
nonanalytic function of a control parameter. Based on the analogy we
established so far, we expect to see a nonanalytic behaviour on
$\mathcal{L}$ when there is a DQPT, since it is a dynamical analogue
of thermodynamic free energy. The rate function is given as:
\begin{eqnarray}\label{rate function}
&& \mathcal{L}(t)=-\lim_{N\rightarrow\infty}\frac{1}{N}\ln\vert\mathcal{G}(t)\vert^2,
\label{eqn:prob. of return amplitude}
\end{eqnarray}
where $N$ is the number of degrees of freedom of the system.

DQPT is intensively studied during the last decade
\cite{heyl2018dynamical,kyaw2020dynamical,PhysRevA.103.012204,PhysRevB.94.064423,PhysRevB.101.014301,2015ARCMP...6..201L}.
In the present work we are interested in the interplay between
topology and DQPTs considering the quantum skyrmion
\cite{PhysRevB.103.L060404}. The scientific community is still
fascinated about topological states of matter even though it has been
over four decades since the discovery of the Quantum Hall state, the
first discovered topological state \cite{Prange_book}. Topologically
distinct states or topological states are those states, which are
classified based on a certain invariant \cite{Altland_book,Wen_book}.
Such states are said to be identical when we can move from one state
to another by applying continuous smooth deformations(deformations
which do not close the bulk energy gap) without changing the value of
the invariant \cite{2011RvMP...83.1057Q}. When the system shows this
kind of resistance to deformation we say that it is topologically
protected. Conventional states, which are earlier believed to be the
same may become topologically distinct. We discover this only when the
accompanying physical behaviour is detected, like in the case of the
Quantum Hall effect.

In the past few decades, scientists discovered many topological
materials and states like topological insulators
\cite{PhysRevLett.61.2015,doi:10.1126/science.1148047}, topological
crystaline insulators \cite{2012NatCo...3..982H, 2012NatPh...8..800T},
topological semi-metals \cite{PhysRevB.76.045302,2018NatPh..14..918S}
etc. This classification is based on the behaviour of the corresponding
band Hamiltonians in the reciprocal state. There are also nontrivial
topological objects determined by their characteristics in real space
such as various types of topological defects in condensed matter
\cite{Mermin_review}.

Magnetic skyrmions are among the most popular types of topological
defects studied now. Skyrmions are particular examples of solitons
that can be informally defined as localized waves with a stable shape
(for more accurate definition and detailed discussion see e.g.
\cite{book_rajaraman}). They are related to
peculiar localized non-colinear magnetic textures within magnetic systems
\cite{piette1995multisolitons,rajaraman1982solitons,skyrme1994non}. Skyrmions
are promised to be potential information carriers for the next
generation of spintronic devices \cite{2020NatRP...2..492B}. New
studies suggest macroscopic skyrmion qubits design suitable for
quantum computing technology controlling the helicity and dynamics of
the skyrmions through electric fields \cite{2021PhRvL.127f7201P}.
These developments make the study of the dynamics of quantum skyrmions demanding.

The formation of magnetic order in spin systems depends
on different factors and competing interactions.
Formation of non-colinear magnetic textures is mainly fueled either by
competing nearest-neighbor ferromagnetic and next nearest-neighbor
antiferromagnetic or asymmetric exchange interactions termed as
Dzyaloshinskii–Moriya interaction (DMI). In most cases, the DMI is the
dominant mechanism forming non-conventional magnetic textures
\cite{belavin1975metastable,barton2020magnetic,schroers1995bogomol,seki2012observation,wilson2014chiral,
  schutte2014magnon,white2014electric,derras2018quantum,haldar2018first,leonov2015multiply,
  psaroudaki2017quantum,van2013magnetic,rohart2016path,samoilenka2017gauged,battye2013isospinning,
  jennings2014broken,tsesses2018optical}.

Recently, the quantum analog of magnetic skyrmions has been suggested
and studied
\cite{PhysRevX.9.041063,2019JPCM...31G5001G,PhysRevB.103.L060404}. However,
contrary to the classical skyrmion the quantum skyrmion is not
topologically stable in a rigorous sense. Qualitatively, it is not
protected with respect to the quantum tunneling to the topologically
trivial vacuum state. At the same time, it presents a quantum spin
state with quite a special character reminiscent of its topologically
protected classical analog. Here we use the words "topological phase"
for the case of quantum skyrmions in this, not completely rigorous but
intuitively clear, sense. To better understand the nature of this
state we studied its real-time dynamical properties.


\begin{figure}[t!]
  \includegraphics[width=1.0\columnwidth]{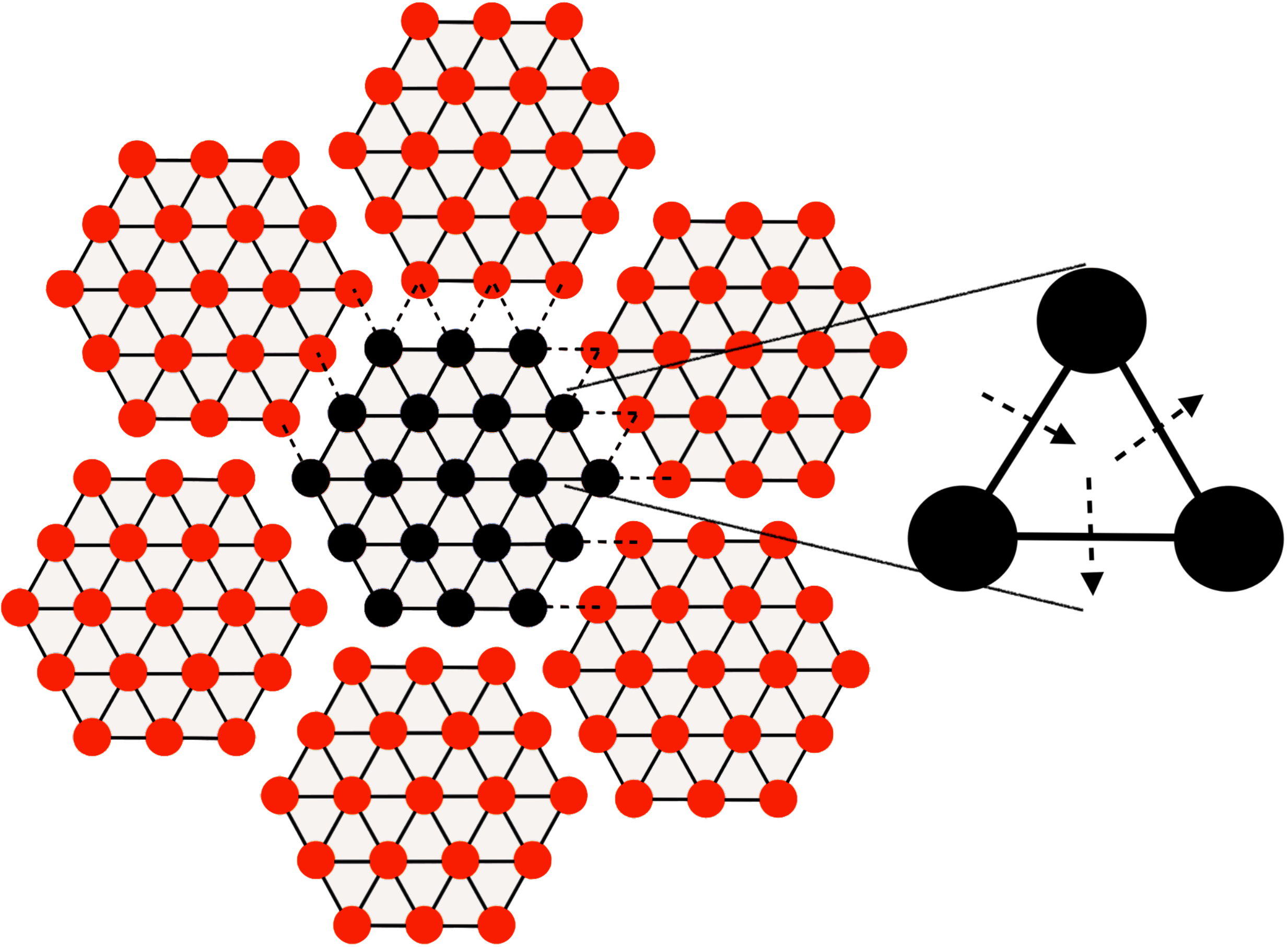}
  \caption{ PBC is applied to the 19 spins central super-cell(black
    unit of the lattice). Super cell is repeated due to the PBC such
    that it forms a larger triangular lattice. The solid bonds show
    bonds within the boundary and the dashed bonds represent PBC. The
    bonds of this lattice can be classified into three based on their
    orientation. Hence the direction of corresponding DMI vectors is
    also different for each one of these categories of bonds. Three
    types of bonds and direction of corresponding DMI vectors(dashed
    arrows) can be seen in the figure inset.} 
\label{fig:system}
\end{figure}

\begin{figure}[ht]
  \includegraphics[width=1.0\columnwidth]{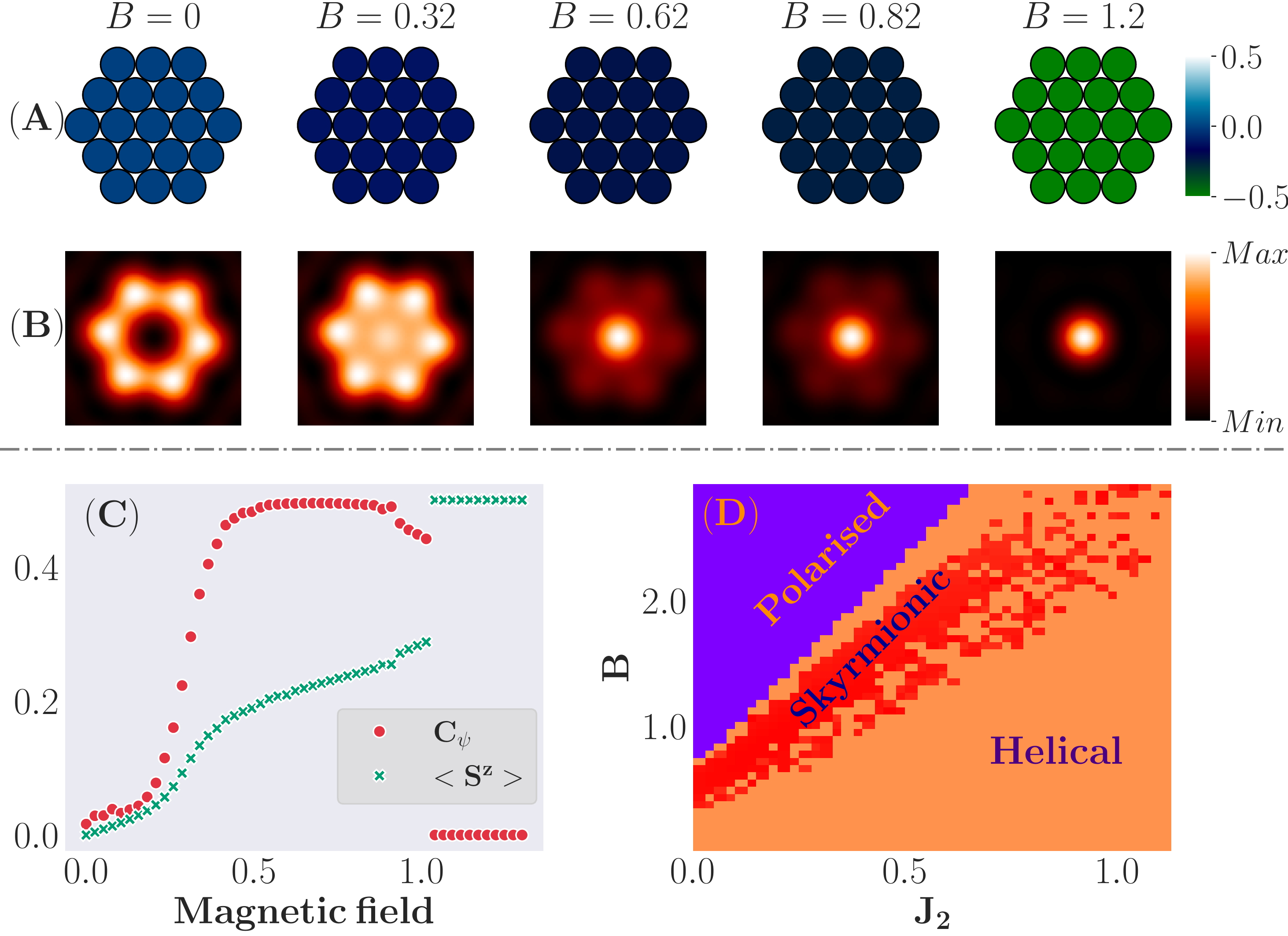}
  \caption{(A) Locations of spins and spin textures of the local
    magnetization of a system at different values of an applied
    magnetic field. Here $D=1, \:J_1=-0.5D, \:J_2=0.1D$. The colors
    quantify expectation values of $\hat S_z$ components of specific
    spins. (B) The Fourier transform of the longitudinal spin
    correlation function for $D=1, \:J_1=-0.5D, \:J_2=0.1D$. The
    observed intensity peaks confirm the formation of non-trivial
    magnetic textures. (C) For $D=1, \:J_1=-0.5D, \:J_2=0.1D$, the
    quantum scalar chirality remains constant between magnetic field
    values of $0.46D$ and $0.86D$. (D) A phase diagram showing
    ferromagnetic, skyrmionic and helical phases using the scalar
    chirality in the $J_2-B$ parametric space.}
\label{fig:skyrmion phase}
\end{figure}

\begin{figure*}[!ht]
\centering
\includegraphics[width=1.0\textwidth]{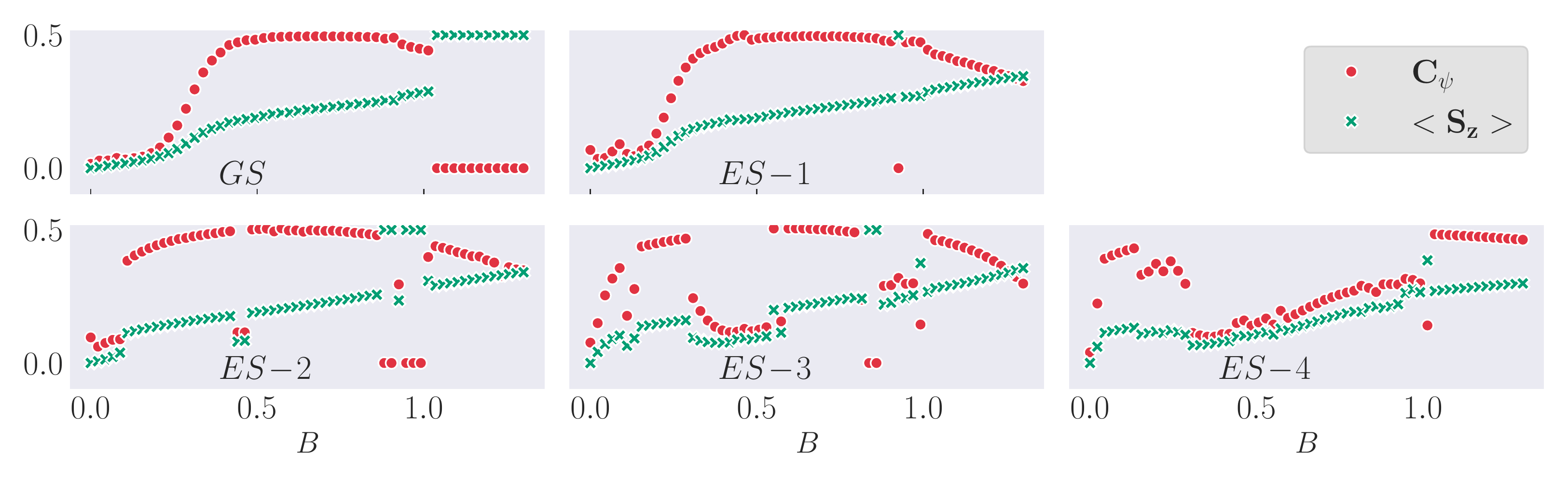}
\caption{Chirality and magnetization vs field for first few eigen
  states for $D=1, \:J_1=-0.5D, \:J_2=0.1D$.  Here in the panel GS
  stands for the ground state, and ES stands for exited states. In GS,
  ES-1, ES-2, and ES-3, we can identify the scalar chirality $C_\psi$
  with the almost constant value (at $C_\psi \sim 0.5$) corresponding
  to a certain range of the magnetic field $B$. For the GS the range
  of $B$ is $ B\sim0.48D$ to $B\sim0.91D$. In ES-1 it is $ B\sim0.48D$
  to $B\sim0.89D$. In ES-2 it is $ B\sim0.48D$ to $B\sim0.85D$. In
  ES-3 it is $ B\sim0.59D$ to $B\sim0.79D$. We notice that for higher
  excited energy levels the range of the plateau is diminished, and at
  ES-4 we completely lost the plateau. An important fact is that
  already for the first excited state ES-1, we observe fluctuation in
  the case of the strong field $ B\sim0.9D$. For higher excited states
  ES-2 and ES-3 fluctuations are enhanced and features of the quantum
  skyrmion state are absent.}
\label{fig:chirality_levels}
\end{figure*}

Unlike skyrmions,
\cite{OchoaTserkovnyak,PhysRevB.98.024423,stepanov2019heisenberg,PhysRevX.9.041063,PhysRevLett.125.227201}
helical magnetic textures do not possess a topological invariant and a
topological protection. Any classification of 2D helical phases is
difficult, even properties of quantum skyrmions are not well explored.
Recent attempts to discover quantum skyrmions show a certain degree of
success. Lohani {\it et.al.} \cite{PhysRevX.9.041063} and Gauyacq {\it
  et.al.} \cite{2019JPCM...31G5001G} could identify magnetization
patterns of a quantum skyrmion but topological protection of skyrmion
phase was not explored. Sotnikov {\it et.al.} used a quantum scalar
chirality to identify a topological
protection\cite{PhysRevB.103.L060404}.
Siegl {\it et.al.} using
topological index and winding parameters, were able to identify a
skyrmion phase and quantify its stability
\cite{2021arXiv211000348S}. All these works could only find a skyrmionic
phase in the ground state.  In the present work, we investigate
the stability of quantum skyrmions in higher excited levels as well.

An important question is the strength of the DM
interaction. DMI can be determined in experiments or
  calculated accurately from first-principles~\cite{Fert2017,Yang2023}.   Our
interest is focused on materials with large DMI:
\cite{PhysRevB.106.024419,PhysRevB.101.184401,PhysRevB.102.241107}.
In recent experimental works H. Yang {\it et.al}
\cite{doi:10.1063/1.5050447} has shown that magnetic films sandwiched
between nonmagnetic layers exhibit DMI with an electrically tunable
strength, {\it i.e.}, \textbf{Co} films sandwiched between nonmagnetic
layers or \textbf{MgO/Fe/Pt}. The value of DMI in such materials
linearly increases with the applied external electric field
$D=D_0+g_{ME}E$, where $D_0$ is the intrinsic DMI part, $g_{ME}$ is
the magnetoelectric coupling and $E$ is the external electric
field. Enhanced DMI can formally reach the order of exchange
interaction for a large electric field. The dynamic control of
intrinsic magnetic interactions by varying the strength of a
high-frequency laser field allows further enhancing of the ratio
between DMI and exchange interaction constants
\cite{stepanov2019heisenberg,PhysRevLett.118.157201,PhysRevLett.115.075301,mikhael1}.
The idea relies on the fact that DMI and exchange interaction are both
based on hopping processes and that time-periodic fields renormalize
the electronic tunneling, leading to the effective rescaled DMI and
exchange constants $D\approx\frac{4t\Delta}{U_{0}-U}$ and
$J\approx\frac{4t^2}{\Omega^2}(U_0-U)$. Here $\Delta$ describes the
Rashba spin-orbit coupling, $t$ denotes the hopping amplitude, $U_0-U$
is the gain in the Coulomb energy due to the electron displacement,
and $\Omega$ is the frequency of the laser field.  Because of the
factor $1/\Omega^2$, the high-frequency laser field can substantially
reduce the rescaled exchange constant to achieve the condition $D>J$.
In order to study Helical-Skyrmionic-Ferromagnetic phases we consider
an array of spins formed in a triangular lattice. The spins are
arranged in such a way that it has a six-fold rotation symmetry (see
Fig.~\ref{fig:system}), also with periodic boundary conditions (PBC)
it possesses translation symmetry.  The Hamiltonian of the system has
the form: 
\begin{eqnarray}
\label{DM-skyrmion}
 \hat H=B\sum\limits_i \hat {\mathbf{S}_i^z}+J_1\sum\limits_{\langle i,j\rangle}\hat {\mathbf{S}_i}\hat {\mathbf{S}_j}+&J_2\sum\limits_{\langle{\langle i,j\rangle}\rangle}\hat {\mathbf{S}_i}\hat {\mathbf{S}_j}\nonumber\\
 +\sum\limits_{i<j}\textbf{D}_{i,j}\left[\hat {\mathbf{S}_i}\times\hat {\mathbf{S}_j}\right],
\end{eqnarray}

where summation in single brackets is taken over the nearest neighbors
and in double brackets over the next-nearest-neighbors, $B$ is an
external magnetic field, and the DMI vector $\textbf{D}$ is aligned
perpendicular to the bond between lattice sites $i$ and $j$, see inset
of Fig.~(\ref{fig:system}). The direction of DMI vectors is chosen to
ensure that the six-fold rotation symmetry holds. We note that in
helical multiferroic insulators, the parameter $D$ is an effect of the
magnetoelectric coupling $g_{ME}$ with an external electric field $E$,
i.e., $D=g_{ME}E$. Thus, the strength of the DMI term can be
controlled externally \cite{PhysRevLett.125.227201}. Here we consider
both $J_1$(a ferromagnetic nearest neighbor exchange) and $J_2$ (an
anti-ferromagnetic next nearest neighbor exchange) interactions along
with the DMI.

An array of spins read along with the above Hamiltonian with a
specific parameter set forms a quantum skyrmion. The nearest-neighbor
ferromagnetic $J_1$ term encourages colinear spin orientation while
DMI term compels non-colinear spin texture. These two competing
interactions form classical skyrmions stabilized by the applied
magnetic field. Below we show that adding even small next nearest
neighbor antiferromagnetic Heisenberg term $J_2$ improves the
stability of quantum skyrmion structures. In general, it is well-known
that the $J_2$ term leads to the spin frustration and formation of
antiferromagnetic classical skyrmions \cite{PhysRevLett.108.017206}
However, quantum skyrmion structures are quite specific as compared to
classical skyrmions. In particular, quantum skyrmions do not possess
continuous magnetic texture and topological charge. Therefore to infer
the quantum skyrmion state, we exploit another tool, such as scalar
chirality.  The energy levels of the system Eq.(\ref{DM-skyrmion})
show a six-fold degeneracy even at very high magnetic field values. We
calculated the expectation values considering maximally mixed state of
these degenerate states with equal probability.

We used spin correlation functions to characterise quantum non-trivial
magnetic structures. In particular, we explore the Fourier transform
of the longitudinal spin correlation function
$G_{\|}(\textbf{q})=\dfrac{1}{N}\sum\limits_{ij}G_{\|}(\textbf{r}_{ij})\exp(-i\textbf{q}\textbf{r}_{ij})$,
where $N$ is the number of spins and the correlation function is given
by
\begin{eqnarray}
\label{correlation functions}
G_{\|}(\textbf{r}_{ij})=\frac{1}{Z}\sum\limits_n\langle n\vert\hat S_i^z\hat S_j^z\vert n\rangle\exp(-\beta E_n).
\end{eqnarray}
where $E_n$s are the eigenvalues of the Hamiltonian of the system and
$r_{ij}$ is the distance between the lattice sites $i$ and $j$.

Topologically protected systems like skyrmions tend to show resistance
to a deformation in its
configuration\cite{2020,doi:10.1080/00018732.2012.663070}. Throughout
this work an applied magnetic field is considered as a cause of
deformation. In Fig.~\ref{fig:skyrmion phase}(A) and
Fig.~\ref{fig:skyrmion phase}(B) panels are arranged in the increasing
order of applied magnetic field from left to
right. Fig.~\ref{fig:skyrmion phase}(A) shows that the local
magnetisation is uniform and increasing magnetic field causes the
local magnetisation to increase in the direction of the field. This
observation fails to identify any magnetic textures in the
system. Fig.~\ref{fig:skyrmion phase}(B), multiple intensity spots
(Bragg peaks) observed for nonzero \textbf{q}(Fourier conjugate of $r$
or wave vector in reciprocal space) confirms formation of the quantum
non-trivial magnetic texture \cite{stepanov2019heisenberg}.  This
quantity could distinguish between helical and ferromagnetic phases
but it fails to identify skyrmionic phase within the helical phase. So
we require another quantifier that can trace out the skyrmion phase
from helical and ferromagnetic phases.

Scalar chirality \cite{PhysRevB.103.L060404}
\begin{eqnarray}
\label{DM-skyrmion_chirality}
&& \textbf{C}_\Psi =\frac{\mathcal{N}}{\pi}\left\langle \hat S_1\left[ \hat S_2\times\hat S_3\right] \right\rangle.
\end{eqnarray}
is considered as a distinguishing property of a helical spin
system. When the chirality has a non-zero value we say that the system
is in a helical phase. It is proposed that chirality can distinguish
quantum skyrmion phase from other phases of the system
\cite{PhysRevB.103.L060404}. 

In this equation $\mathcal{N}$ is the number of non-overlapping elementary triangular
patches covering the lattice. Three adjacent spins form a
patch. The scalar chirality for any of these three adjacent spin
combinations is the same, because of the translational and
rotational symmetries of the lattice.

In Fig.~\ref{fig:skyrmion phase}(C) from $B=0D$ to $B=0.46D$ chirality
increases almost steadily. In this region no two deformed states have
same chirality value, therefore, all those states are topologically
non-identical. From $B=0.46D$ till $B=0.86D$ the plateau of scalar
chirality implies that all the states in this region have a common
topological invariant and we call them topologically identical phase
or simply topological phase. After crossing $B=0.86D$ the system
briefly falls back to a helical phase. It is represented by a dip in
chirality. Then around $B=1.0D$ the system goes to the trivial
ferromagnetic phase indicated by zeros of scalar chirality.

In Fig.~\ref{fig:skyrmion phase}(C) The magnetisation
graph is telling us that for the region where we have plateau in
chirality the magnetisation is not same for any two states {\it i.e.},
the system is in fact undergoing deformation during the constant
chirality plateau also.

The phase diagram (Fig. \ref{fig:skyrmion phase}(D)) shows that for
non zero $J_2$ the invariance of scalar chirality extends for larger
values of the applied magnetic field compared to $J_2=0$. The skyrmion
phase is embedded into the helical phase. For a larger value of the
interaction parameter $J_2$, the system retains the skyrmion phase for
a longer range of applied magnetic field. We studied the scalar
chirality not only in the ground state but also in several excited
states. We see that the skyrmion state survives in first and second
excited states as well as a certain degree of topological invariance
can be seen in higher excited states, see
Fig.~\ref{fig:chirality_levels}. When $J_2=0D$ we arrive at the
results from O.\,M. Sotnikov {\it
  et. al}. \cite{PhysRevB.103.L060404}. For higher values of
$J_2$($J_2>0.35D$) the plateau gets distorted.

\begin{figure}[t]
  \includegraphics[width=1.0\columnwidth]{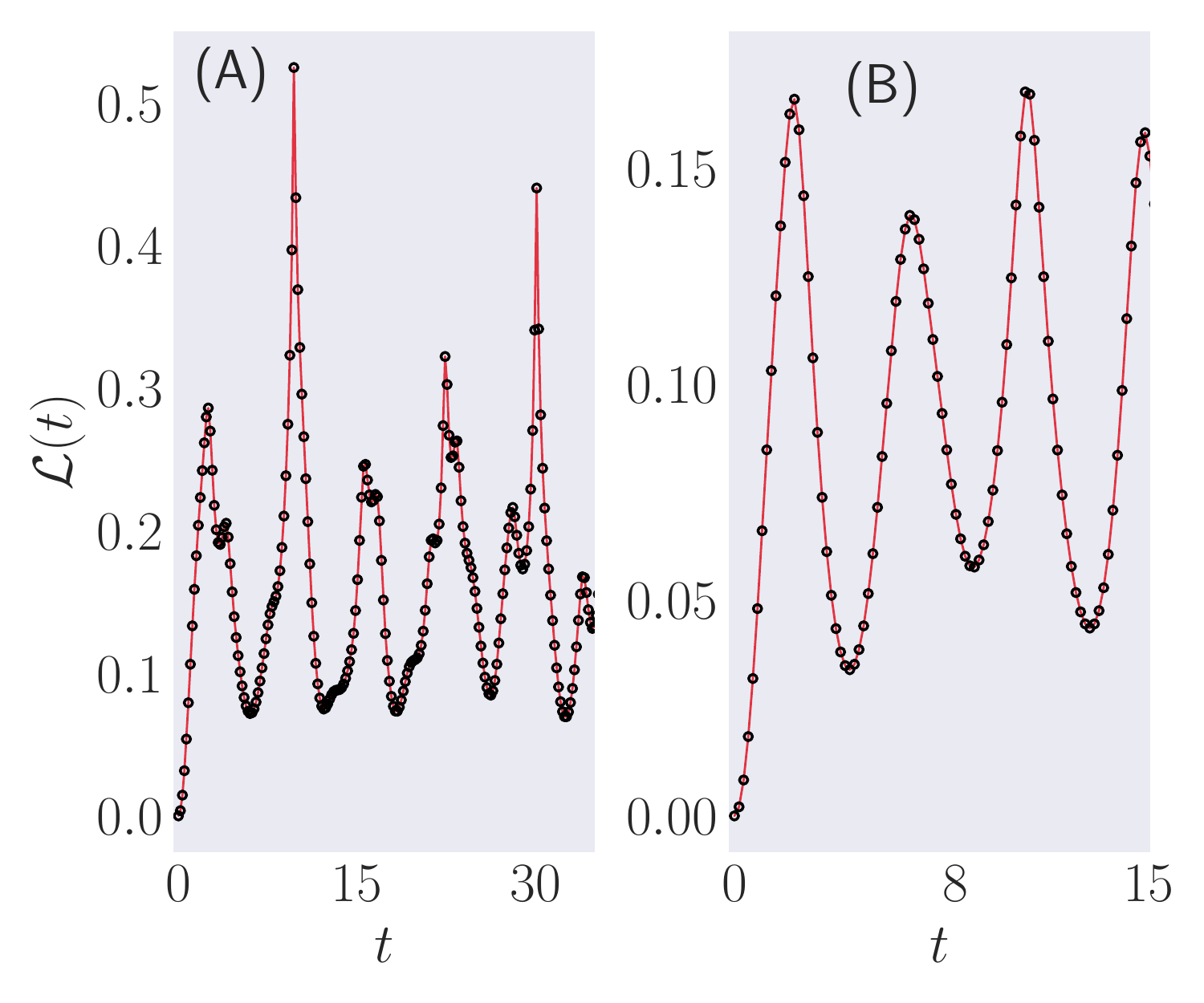}
  \caption{Quench protocol is applied on ground state initiated with
    $D=1$, $J_1=-0.5D$, $J_2=0.1D$, $B=0.85D$. (A) Time evolution of
    rate function for skyrmionic to polar phase transition. At $t=0$
    DMI is turned off (set $D=0$). The peak of rate function at
    $t= 9.71,\:30.1\: etc.\ $ shows nonanalytic behaviour. (B) Time
    evolution of rate function for skyrmionic to Helical phase
    transition. At $t=0$ the applied field, $B$ is turned off. Here we
    do not find any nonanalyticity.}
\label{fig:DQPTs}
\end{figure}

Here we
show that an important information about helical and quantum skyrmion
phases can be obtained from the analyses of topological DQPTs. Namely,
when quenching the system from a quantum skyrmion to a
topologically trivial phase, we observe a characteristic signature
of a DQPT as a nonanalyticity in the rate function.  On the other hand,
a quench between a skyrmion and a helix does
not lead to a DQPT.

Let the system be prepared in the ground state $\vert\psi_0\rangle$ of
the Hamiltonian $\hat H_0=\hat H(\lambda_0)$. At $t=0$ the parameter
$\lambda$ is quenched to a new value $\hat H(\lambda_f)$ and the initial
wave function is evolved to a new state
$\vert\psi(t)\rangle=e^{-i\hat H t}\vert\psi_0\rangle$ under this new
Hamiltonian. In order to describe the DQPT we study the Loschmidt
amplitude (return amplitude) and the rate function, in the
thermodynamic limit
\cite{PhysRevLett.113.205701,PhysRevLett.118.180601,PhysRevLett.121.130603,PhysRevLett.124.043001,PhysRevLett.124.250601,PhysRevResearch.2.033259}.

Our primary interest concerns DQPTs between a quantum skyrmion phase
and a trivial phase. The quench protocol applied on a topological
phase to a trivial (skyrmion to ferromagnet) phase produced a
nonanalytic behaviour of the rate function with respect to time.  The
result is plotted in Fig.~\ref{fig:DQPTs}(A). On the other hand, when
switching between a skyrmionic phase to a helical phase, we do not
observe any DQPT. We can confirm this from the lack of nonanalytic
behaviour of the rate function when plotted against time in
Fig.~\ref{fig:DQPTs}(B). Here we looked for nonanalyticity with
different values of the parameter $B$ but we could not find any. The
presence of nonanalyticity in the former case can be explained by a
sharp and discontinuous transition of $\boldsymbol{C_{\psi}}$ from
skyrmionic to trivial phase in Fig.~\ref{fig:skyrmion
  phase}(C). Similarly the lack of nonanalyticity in the later case is
due to the smooth and continuous transition of $\boldsymbol{C_{\psi}}$
from helical to skyrmionic phase. From Fig.~\ref{fig:DQPTs}(A) the
critical exponent $\alpha = 0.7020 \pm 0.0233$ around $t_c =
9.66$. The value turned out to be very close to this for other
critical points as well (see supplementary methods for details section
\ref{Supplementary}, see, also, references \cite{PhysRevB.104.115159,PhysRevLett.115.140602,PhysRevLett.126.200602} therein). Further study is required with larger system
sizes($L=3i(i+1)+1, \: i=3, \, 4, \,5, \,...$) in order to comment on
the universality of the critical exponent that we calculated.

We have computational limitations to analyzing finite-size effects and
artifacts of a particular quantum skyrmion state in our
system. However, we performed the study of the finite-size effects
using another quantum skyrmion state obtained for a slightly different
model in \cite{2021arXiv211000348S}. The results of the calculations
are shown in section \ref{Supplementary}. We see the same trend for this case
also, {\it i. e.}, non-analytic singularities in the rate function
during the dynamical quantum phase transition between the skyrmion and
FM phases. Thus obtained results are pretty universal and apply to any
quantum skyrmion. The reason for the universal effect is the
orthogonality of quantum skyrmion and FM states. This argument is
valid for any quantum skyrmion independently of its size.

Apart from this, we achieved a significant improvement of the
topological phase stability with our model compared to the previous
works
\cite{PhysRevB.103.L060404,PhysRevX.9.041063,2021arXiv211000348S}. We
note that within a skyrmionic phase a larger $J_2$ value gives a
topological phase protection against a larger range of applied
magnetic fields. At high values of $J_2$ the topological invariance is
destroyed. This result tells us that the key for tunability and
improved stability of quantum skyrmions can be the interaction
parameter, this may be useful when choosing the material to realise
skyrmions for experiments. This kind of model is realized in {\bf
  Pd/Fe/Ir(111)} system with {\bf Co} surrounded edges
\cite{Spethmann_2022}. Also nanoscale skyrmions are reported at room
temperature with large DMI interaction in {\bf Ir/CoFeB/MgO} systems
\cite{CHEN2021100618}.

In conclusion, the quantum skyrmion model proposed above shows
significant improvement in topological protection. In the said model
we identified a robust DQPT when quench protocol is applied from a
skyrmionic state to a ferromagnetic state. Robust DQPTs accompany many
interesting properties.  An interesting direction is to look for the
connection between entanglement dynamics and DQPTs in a skyrmionic
phase. Certain systems have reported to show an enhanced entanglement
entropy around critical points of DQPTs \cite{PhysRevLett.119.080501}.

\section*{Acknowledgement}

We would like to sincerely thank O. M. Sotnikov for discussion related
to the initial stage of this work. We acknowledge National
Supercomputing Mission (NSM) for providing computing resources of
‘PARAM Shivay’ at Indian Institute of Technology (BHU), Varanasi,
which is implemented by C-DAC and supported by the Ministry of
Electronics and Information Technology (MeitY) and Department of
Science and Technology (DST), Government of India.  A.E. acknowledges
funding by Fonds zur Förderung der wissenschaftlichen Forschung (FWF)
grant I 5384. The work of MIK is supported by European Research
Council via Synergy Grant 854843 - FASTCORR. SKM acknowledges Science
and Engineering Research Board, Department of Science and Technology,
India for support under Core Research Grant CRG/2021/007095.

\section{Supplemental material to ``Topological dynamical quantum phase transition in a quantum
skyrmion phase"}
\label{Supplementary}
\maketitle
\section*{Calculation of critical exponent \texorpdfstring{$\mathbf{\alpha}$}{ }}
We adopted a method discussed by Trapin {\it et al} \cite{PhysRevB.104.115159} to calculate the critical exponent of DQPT. For a one dimensional system Trapin {\it et al} report $\alpha =0.1264(2)$ which is less than the value defined by universality class of the problem. They call it as an unconventional critical exponent.

It is a well known fact that the critical exponents depend on factors like dimension of the system, range of the interaction and spin dimension.
\begin{figure}[!ht]
  \centering
\includegraphics[width=0.8\columnwidth]{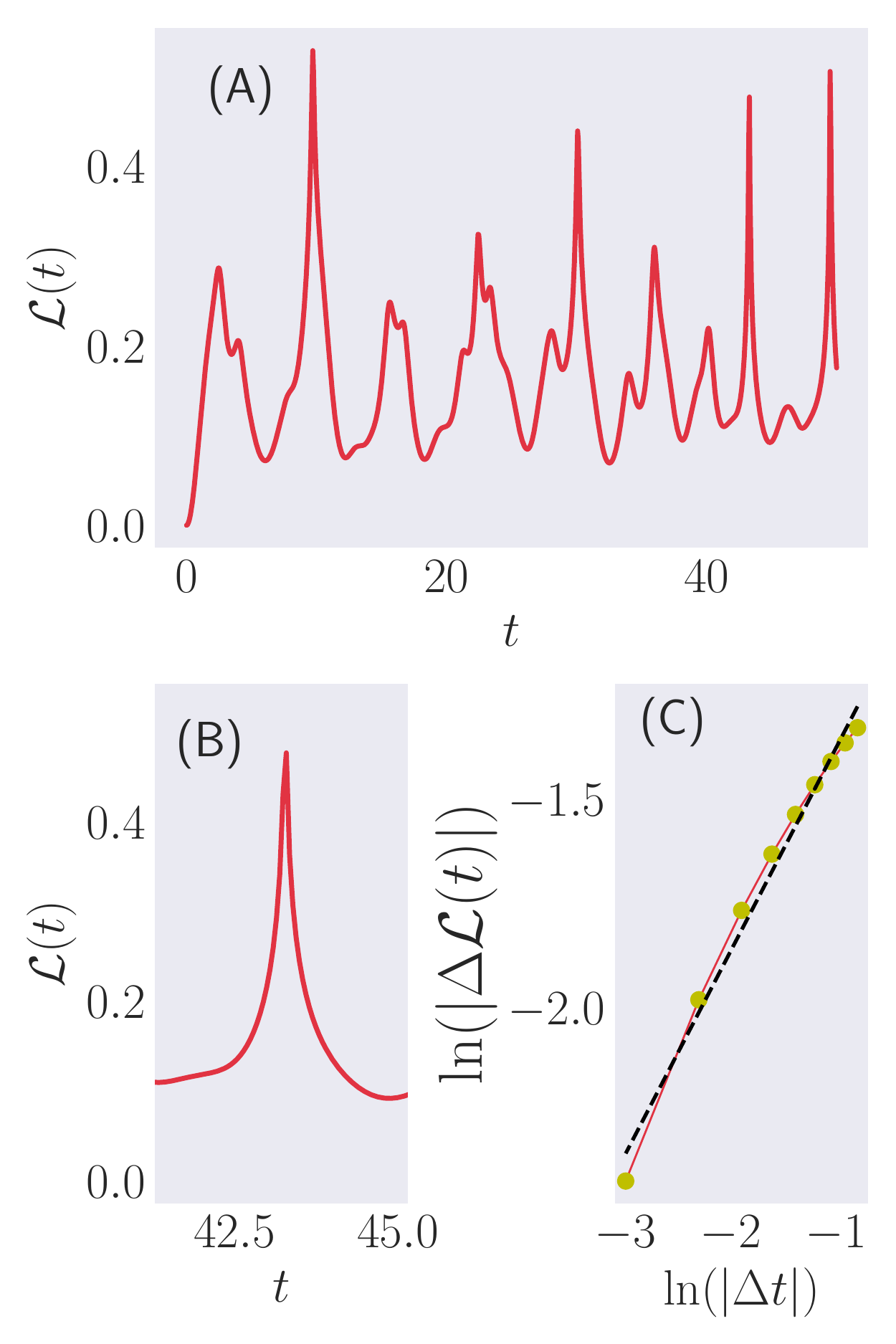}
  \caption{{\bf How to calculate critical exponent associated to a nonanalyticity.} (A) Rate function showing nonanalyticity. (B) Zoomed in on the first nonanalytic point at $t\sim43.24$. (C) $\ln (\vert \Delta \mathcal{L} \vert)$ vs. $\ln (\vert \Delta t \vert)$: Red line with green dots is observed data, black dashed line is the straight line approximation.}
\label{Fig:critical}
\end{figure}

\begin{figure*}[!ht]
  \centering

\begin{tabular}{cc}
    \includegraphics[width=0.8\columnwidth]{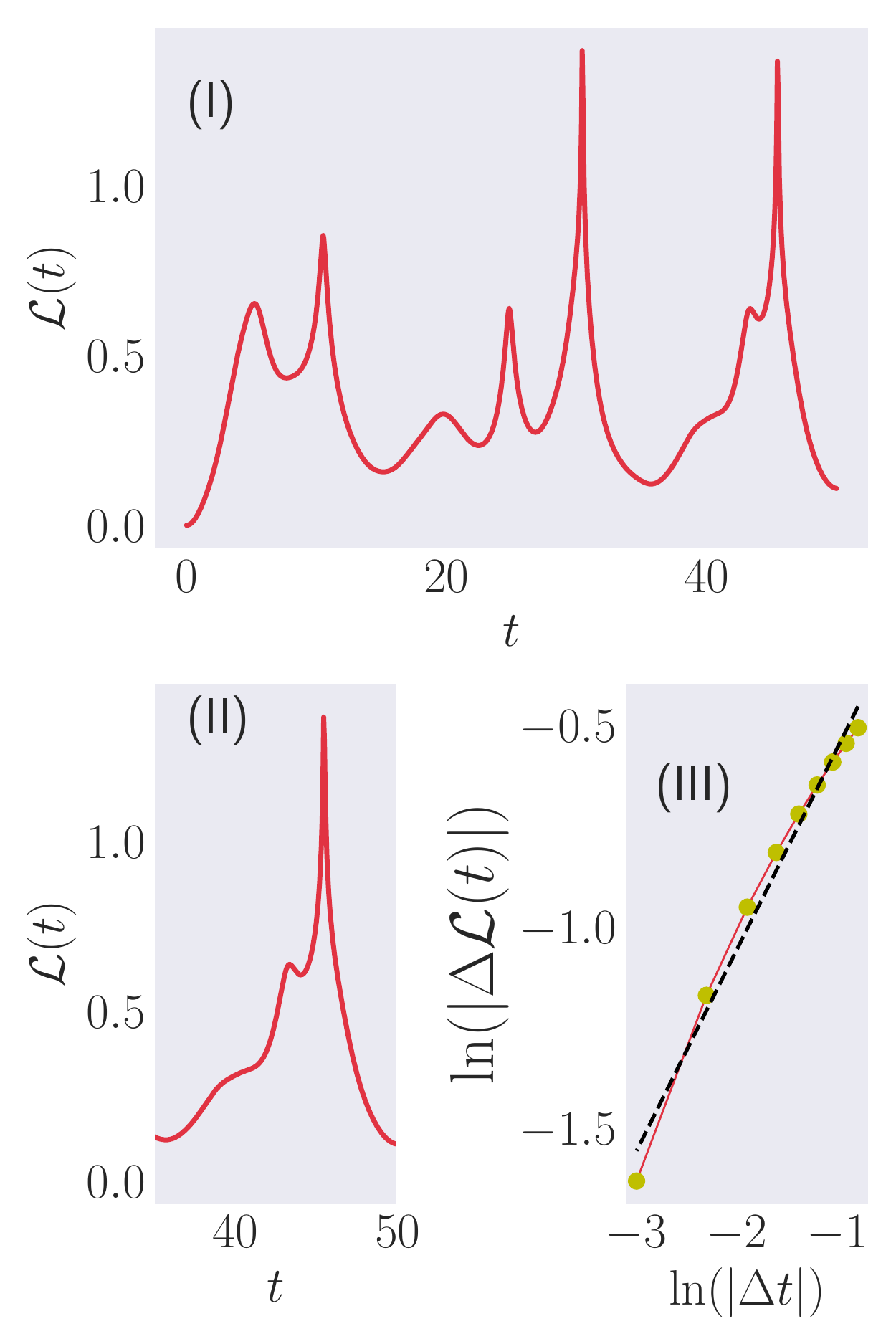} &  \includegraphics[width=0.8\columnwidth]{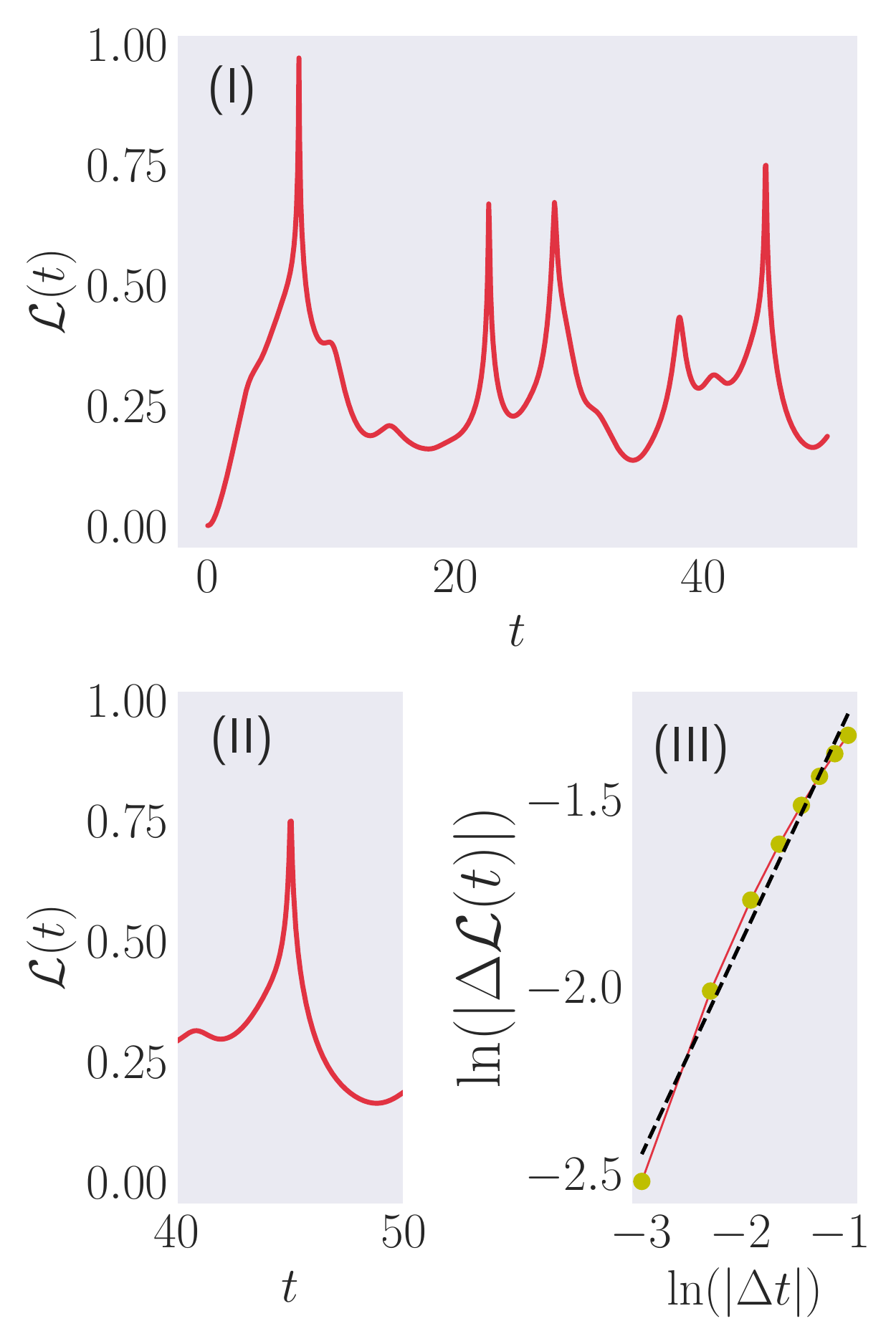}
 \\

 (A) & (B)
\end{tabular}
  \caption{{\bf (A) Critical exponent in DQPT of a $3\times3$ square lattice.} (I) Rate function showing nonanalyticity. (II) Zoomed in on a nonanalytic point at $t\sim45.44$. (III) $\ln (\vert \Delta \mathcal{L} \vert)$ vs. $\ln (\vert \Delta t \vert)$: Red line with green dots is observed data, black dashed line is the straight line approximation. {\bf (B) Critical exponent in DQPT of a $4\times4$ square lattice.} (I) Rate function showing nonanalyticity. (II) Zoomed in on a nonanalytic point at $t\sim45$. (III) $\ln (\vert \Delta \mathcal{L} \vert)$ vs. $\ln (\vert \Delta t \vert)$: Red line with green dots is observed data, black dashed line is the straight line approximation.}
\label{Fig:critical_square}
\end{figure*}

Contrary to the system we have in this manuscript, Trapin {\it et al} only consider a one dimensional chain with nearest-neighbour interaction, this can cause a difference in the value of $\alpha$ that we calculate. Attempts were made to find universality and scaling for DQPTs for one and two dimensional systems \cite{PhysRevLett.115.140602}. For 1d system  the exponent $\alpha$ is reported to be 1. 
A recent work by Bandyopadhyay {\it et al.} \cite{PhysRevLett.126.200602} outlines a protocol based on out of time order correlations(OTOC) and string-like observables to experimentally determine the critical exponents of DQPT. Using  the method they show a universal scaling critical exponent for one dimensional systems to be $\alpha=1$. But for 2D systems the analysis in \cite{PhysRevLett.115.140602} could not find a conclusive value for $\alpha$ in general.
Nonetheless we may still calculate $\alpha$ for our system using the method discussed by Trapin {\it et al}.. Let  $t-t_c$ denote a nonanalytic point. We can approximate the rate function $\mathcal{L}(t)$ around the nonanalyticity as follows,
\begin{equation}
    \vert \mathcal{L}(t) - \mathcal{L}_c \vert \sim \vert t-t_c \vert ^ \alpha.
    \label{eqn:critical}
\end{equation}
Taking logarithm on both sides of Eq.~(\ref{eqn:critical}) gives us the equation of a straight line with slope $\alpha$.

Consider $\Delta t=t-t_c$ and $\Delta \mathcal{L}=\mathcal{L}(t)-\mathcal{L}_c$. We plot $\ln (\vert \Delta \mathcal{L} \vert)$ versus $\ln (\vert \Delta t\vert)$ and fit it to a straight line that best fits. The slope of the straight line is the exponent $\alpha$. Let us consider the nonanalytic point in Fig.~\ref{Fig:critical}(A) which is around $t_c = 43.24$. This region is highlighted in Fig.~\ref{Fig:critical}(B). Now if we plot $\ln (\vert \Delta \mathcal{L} \vert)$ versus $\ln (\vert \Delta t\vert)$ and fit a straight line as shown in Fig.~\ref{Fig:critical}(C), the exponent $\alpha$ at $t_c=43.24$ is determined to be $0.4883 \pm 0.0218$ with this method. The critical exponent $\alpha$ turned out to be similar for other $t_c$ values as well, {\it e. g.,} $\alpha = 0.7020 \pm 0.0233$ for $t_c= 9.66$.

\textcolor{black}{Since we have computational limitations in our model to look for different system sizes, to accommodate for the finite size effect we consider the skyrmion model proposed by Siegl {\it et.al} \cite{2021arXiv211000348S}.} 

\textcolor{black}{Consider an $n \times n$ lattice of quantum spin-1/2's coupled to classical ferromagnetic control fields at its boundary. The Hamiltonian of the system is given as,}

\begin{eqnarray}
\label{pia_Hamiltonian}
 \hat H=&-J\sum\limits_{\langle i,j\rangle}(S_i^xS_j^x + S_i^yS_j^y)-\Delta \sum\limits_{\langle i,j\rangle}S_i^zS_j^z\nonumber\\
 &-D\sum\limits_{\langle i,j\rangle}(\mathbf{u_{ij}}\times \hat{z}).\left(\mathbf{S}_i\times\mathbf{S}_j\right),
\end{eqnarray}
\textcolor{black}{with ferromagnetic exchange constant $J>0$, axial Heisenberg anisotropy $\Delta>0$, and the strength of DMI interaction $D$. $\mathbf{u_{ij}}$ is a vector pointing from $\mathbf{S_i}$ to $\mathbf{S_j}$. $\mathbf{\hat{S}_i}=(S_i^x,S_i^y,S_i^z)$ is a vector of spin operators for spins within the $n \times n$ lattice. A layer of classical ferromagnetic spins $S_i = \frac{\hbar}{2}\hat{z}$ resides outside the lattice.}

\textcolor{black}{Further in order to define a skyrmion we utilize the following quantities.}

\begin{equation}
\begin{rcases}
  Q \\
   C
 \end{rcases} = \frac{1}{2\pi} \sum \limits_{\sigma} \tan^{-1}\left(\frac{ \mathbf{n}_i(\mathbf{n}_j \times \mathbf{n}_k)}
 {1+ \mathbf{n}_i \mathbf{n}_j + \mathbf{n}_i \mathbf{n}_k \mathbf{n}_k\mathbf{n}_j}\right), 
\end{equation}
\textcolor{black}{ where the sum runs over all the elementary triangles formed by nearest-neighbour lattice sites $i,$ $j$, $k$ which include the classical ferromagnetic boundary sites as well and no two triangles overlap. The magnitude of the winding parameter $Q$ quantifies the stability of the skyrmion. It is computed with $\mathbf{n_i} = 2 \langle \mathbf{S}_i \rangle \hbar$, where $\langle \mathbf{S}_i \rangle = (\langle S_i^x \rangle, \langle S_i^y \rangle, \langle S_i^z \rangle)$ is the classical magnetic moment or spin expectation value.
    The topological index $C$ takes $\mathbf{n_i} =  \langle \mathbf{S}_i \rangle / 
 |\langle \mathbf{S}_i \rangle|$. $C = \pm 1$ for quantum skyrmions.}

\textcolor{black}{For a $3 \times 3$ lattice when $\Delta=0.5J$ and $D=2J$ we identified a quantum skyrmion phase with $C = -1$ and $Q=-0.981$. We note that parameters $\Delta=0.5J$ and $D=0$ give a quantum ferromagnet with $C=0$ and $Q=0$. Fig.~ \ref{Fig:critical_square}(A) shows the rate function and the calculation of critical exponent for this case. The critical exponent in this case is calculated to be $0.5020 \pm 0.0240$.
Similar observations were done with $4 \times 4$ lattice case also, where parameters $\Delta=0.5J$ and $D=2J$ get a skyrmion phase with $C = -1$ and $Q=-0.981$. $D=0$ gives a ferromagnetic ground state with $C=0$ and $Q=0$. We initiate the system from the skyrmion phase and at $t=0$ we quench the system by setting $D=0$. The results are summarised in Fig.~ \ref{Fig:critical_square}(B).
 The critical exponent in this case is found to be  $0.5667 \pm 0.02878$.}

Next, we investigate DQPT on another 2d model considering a triangular lattice within a circle with ferromagnetic classical boundaries having the following Hamiltonian,

\begin{eqnarray}
\label{triangular_Hamiltonian}
 \hat H=&-J_1\sum\limits_{\langle i,j\rangle}(S_i^xS_j^x + S_i^yS_j^y)-K \sum\limits_{\langle i,j\rangle}S_i^zS_j^z\nonumber\\
 &+\mathbf{D_1}\sum\limits_{\langle i,j\rangle}\left(\mathbf{S}_i\times\mathbf{S}_j\right),
\end{eqnarray}

\begin{figure}[!ht]
  \centering
\includegraphics[width=0.8\columnwidth]{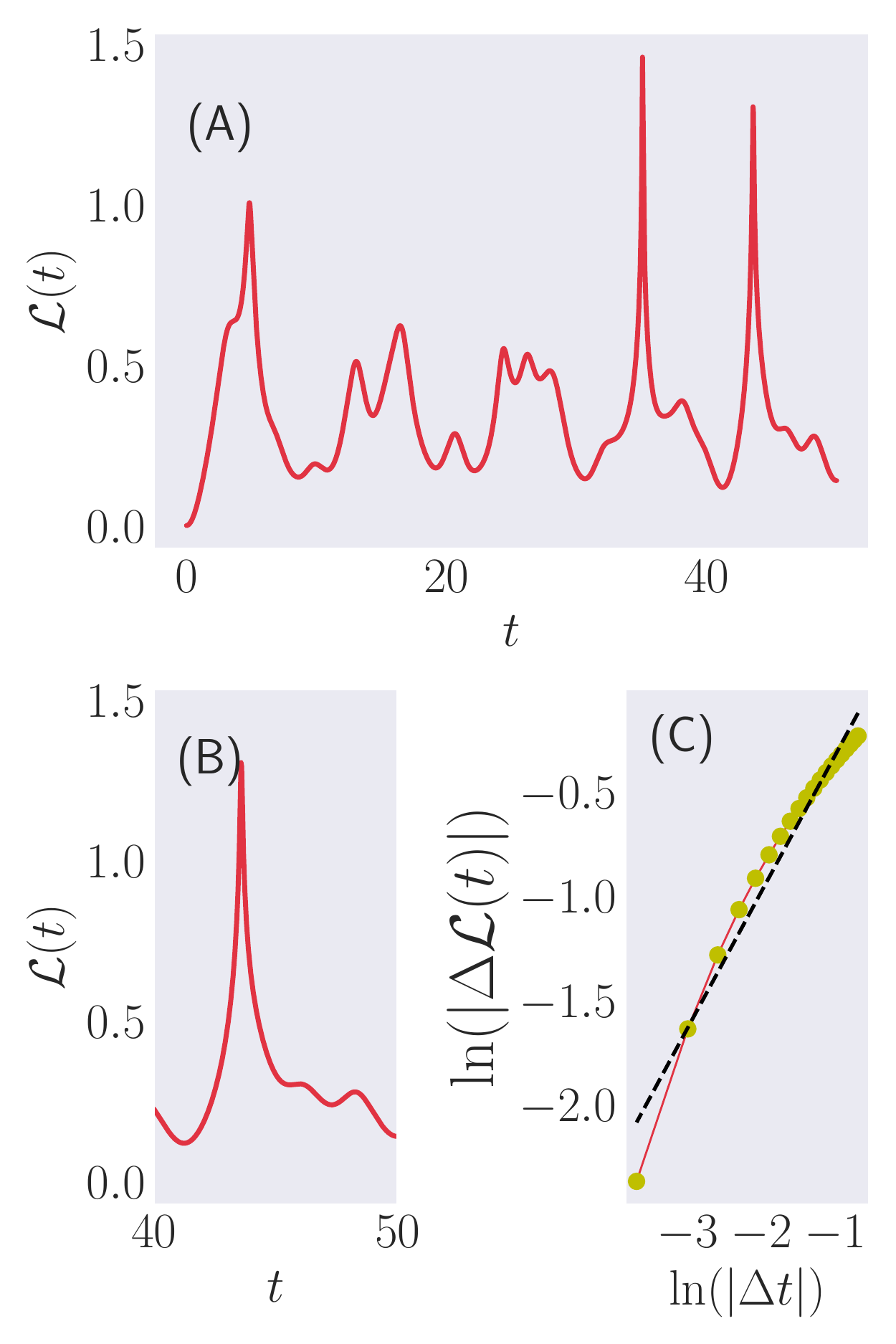}
  \caption{{\bf Nonanalytic rate function and the calculation of exponent for a 7-spin triangular lattice.} (A) Rate function is showing nonanalyticity. (B) Zoomed in on nonanalytic point at $t\sim43.58$. (C) $\ln (\vert \Delta \mathcal{L} \vert)$ vs. $\ln (\vert \Delta t \vert)$: Red line with green dots is observed data, black dashed line is the straight line approximation.}
\label{Fig:critical_triangular}
\end{figure}

\textcolor{black}{Here $J_1 >0$ is the nearest-neighbour ferromagnetic coupling constant, $K >0 $ is the axial Heisenberg anisotropy coefficient and $D_1 > 0$ is the strength of the DMI vector $\mathbf{D_1}$. The direction of DMI is the same as the one shown in Fig.~\ref{fig:system} . $\mathbf{\hat{S}_i}=(S_i^x,S_i^y,S_i^z)$ is a vector of spin operators for spins within the circular triangular lattice. A layer of classical ferromagnetic spins $S_i = \frac{\hbar}{2}\hat{z}$ resides outside the lattice. We make use of the quantities $Q$ and $C$ to identify quantum skyrmions here also.}

\textcolor{black}{In a 7-spin triangular lattice case with circular boundary for $J_1 = 1$, $K=0.61J_1$ and $D_1=2J_1$ we see a skyrmion phase with $C = 1$ and $Q = 0.993$. In the same configuration $D_1 = 0$ gives a ferromagnetic phase with $C = 0$ and $Q = 0$. We initiated the system at the skyrmion phase and quenched it to a ferromagnetic phase by setting $D_1=0$ at $t=0$. The resulting rate function is shown in Fig.~ \ref{Fig:critical_triangular}. The critical exponent calculated is $0.6130 \pm 0.0191$ at $t_c = 43.58$.}

\begin{figure}[!ht]
  \centering
\includegraphics[width=1\columnwidth]{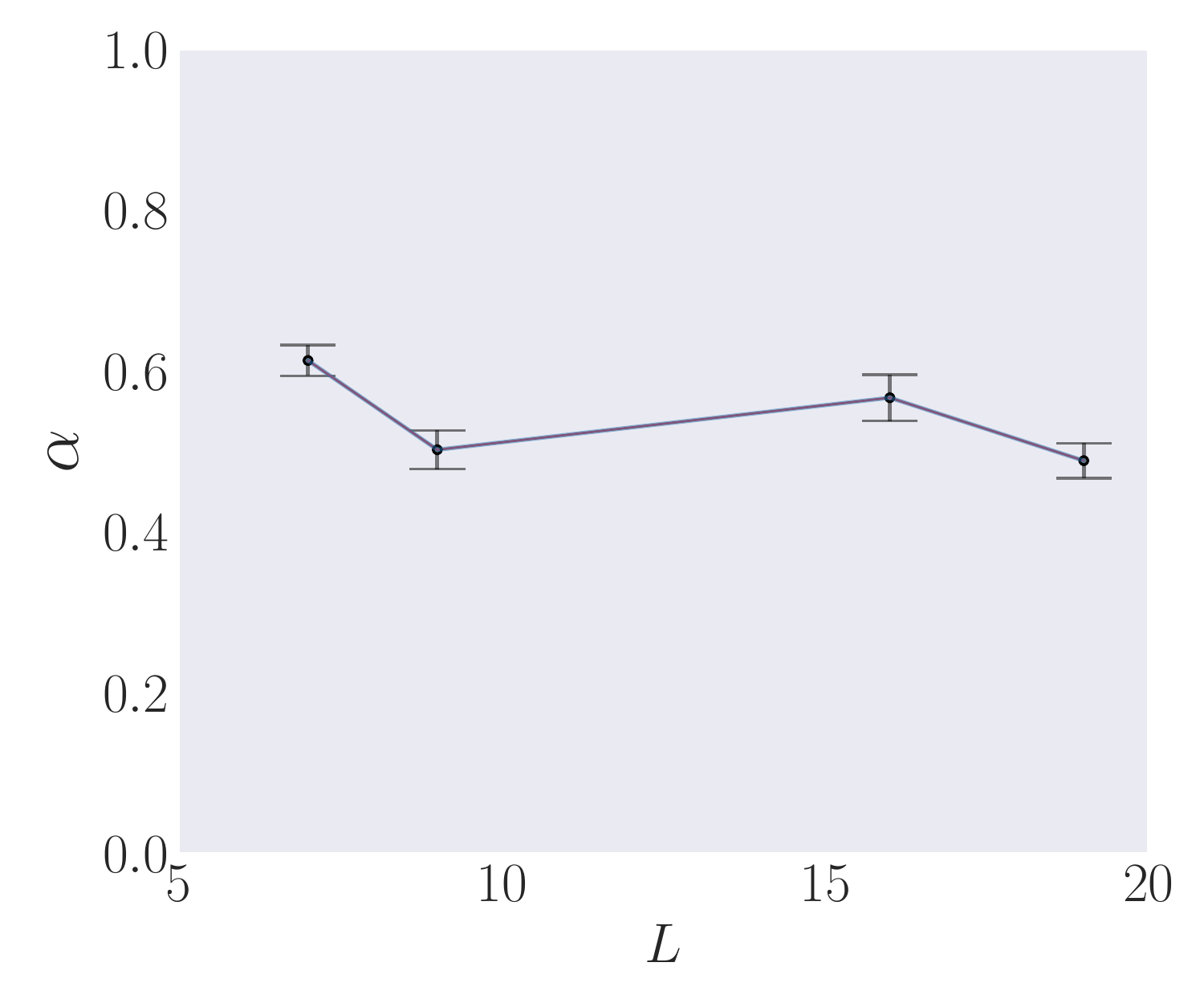}
  \caption{ Critical exponents $\alpha$ in DQPT of three similar 2d models vs. system size $L$ around the same time domain $43.24<t_c<45.44$. All the cases show a universal exponent around $0.5$.}
\label{Fig:finite_size}
\end{figure}

\textcolor{black}{We observed DQPT in various systems with different geometry and interactions around the same time domain $43.24<t_c<45.44$ with system sizes 7, 9, 16, 19 and the critical exponents for respective sizes are $0.6130 \pm 0.0191$, $0.5020 \pm 0.0240$, $0.5667 \pm 0.02878$ and $0.4883 \pm 0.0218$. This is shown in Fig.~ \ref{Fig:finite_size}. In the all cases we find the exponents around 0.5, which hints a universality in skyrmion to ferromagnetic state DQPT. }

\bibliographystyle{apsrev4-1}

\bibliography{references}

\end{document}